\documentclass[12pt]{article}

\usepackage{graphicx}
\usepackage{amsfonts}
\usepackage{amssymb,amsmath,amscd}
\def\+{{+\!\!\!+}}

\def\d{\partial}

\def\pmb#1{\setbox0=\hbox{#1}%
\kern.0em\copy0\kern-\wd0 
\kern-.04em\copy0\kern-\wd0 
\kern.08em\copy0\kern-\wd0 
\kern-.04em\raise.0433em\box0 }         
 
\def\rank{\textstyle{\rm{rank}}} 
 
\newcommand{\nc}{\newcommand} 
\nc{\beq}{\begin{equation}} 
\nc{\eeq}[1]{\label{#1}\end{equation}} 
\nc{\ber}{\begin{eqnarray}} 
\nc{\eer}[1]{\label{#1}\end{eqnarray}} 
\nc{\pek}[1]{\cite{#1}} 
\nc{\enr}[1]{(\ref{#1})} 
\nc{\kal}[1]{{\cal{#1}}} 
\nc{\dott}{\;\cdot\;} 

\newtheorem{Thm}{Theorem}[section]

\newtheorem{Def}[Thm]{Definition}
\newtheorem{Exa}[Thm]{Example}

\newtheorem{cor}[Thm]{Corollary}
\newtheorem{pro}[Thm]{Proposition}

\def\0 {\nonumber}

\begin{document} 
\setcounter{page}{0}
\newcommand{\inv}[1]{{#1}^{-1}} 
\renewcommand{\theequation}{\thesection.\arabic{equation}} 
\newcommand{\be}{\begin{equation}} 
\newcommand{\ee}{\end{equation}} 
\newcommand{\bea}{\begin{eqnarray}} 
\newcommand{\eea}{\end{eqnarray}} 
\newcommand{\re}[1]{(\ref{#1})} 
\newcommand{\qv}{\quad ,} 
\newcommand{\qp}{\quad .} 

\thispagestyle{empty}
\begin{flushright} \small
UUITP-06/06 \\
\end{flushright}
\smallskip
\begin{center} \LARGE
{\bf Lectures on Generalized Complex Geometry and Supersymmetry}
 \\[12mm] \normalsize
{\large\bf Maxim Zabzine} \\[8mm]
 {\small\it
Department of Theoretical Physics 
Uppsala University, \\ Box 803, SE-751 08 Uppsala, Sweden \\
~\\
}
\end{center}
\vspace{10mm}
\centerline{\bfseries Abstract} \bigskip
These are the lecture notes from the 26th  Winter School "Geometry and Physics", 
    Czech Republic, Srni, January 14 - 21, 2006.  These lectures are an introduction 
     into the realm of generalized geometry based on the tangent plus the cotangent bundle.
     In particular  we discuss the relation of this geometry to physics, namely to 
      two-dimensional  field theories. We explain in detail the relation between 
       generalized complex geometry and supersymmetry.  We briefly review 
        the generalized K\"ahler and generalized Calabi-Yau manifolds and explain their
         appearance in physics. 
         
\vspace{2cm}
{\bf to my wife, Natasha Zabzina with love}

\vspace{2cm}
The lecture notes are published in\\ Archivum mathematicum (Supplement) {\bf 42} (2006) 119-146.

\noindent  

\eject

\section*{Introduction}

These are the notes for the lectures presented at the 26th Winter School "Geometry and 
 Physics", Srni, Czech Republic, January 14-21, 2006.  
  The principal aim in these lectures has been to present, in a manner intelligible to both 
  physicists and mathematicians, the basic facts about the generalized complex geometry 
   and its relevance to string theory.   Obviously, given the constraints of time, the discussion of many subjects is somewhat   abbreviated. 
   
  In \cite{hitchinCY}  Nigel Hitchin introduced  the notion of generalized complex structure and generalized  Calabi-Yau manifold.  The essential idea is to take a manifold $M$ and replace the 
   tangent bundle $TM$ by $TM \oplus T^*M$, the tangent plus the cotangent bundle.  The generalized 
    complex structure is a unification of symplectic and complex geometries and is the complex
     analog of a Dirac structure, a concept introduced 
      by Courant and Weinstein \cite{courantw}, \cite{courant}. 
     These mathematical structures can be mapped into string theory. In a sense  they can be 
    derived and motivated from certain aspects of string theory.  The main goal of these lectures is 
     to show the appearance of generalized geometry in string theory. The subject is still in the 
      progress and some issues remain unresolved. In an effort to make a  self-consistent presentation 
       we  choose to concentrate on Hamiltonian aspects of the world-sheet theory and we leave aside
        other aspects which are equally important.  
      
  The lectures are organized as follows. In Lecture 1 we introduce the relevant mathematical 
   concepts such as Lie algebroid, Dirac structure and generalized complex structure. 
    In the next Lecture we explain the appearance of these structures in string theory, in particular
     from the world-sheet point of view. We choose the Hamiltonian formalism as most natural
      for the present purpose. In the  last Lecture we review more advanced topics, such as
      generalized K\"ahler and generalized Calabi-Yau manifolds. 
       We briefly comment on their appearance in string theory. 

Let us make a comment on notation. Quite often we use the same letter for a bundle morphism 
 and a corresponding map between the spaces of sections. Hopefully it will not irritate the mathematicians 
  and will not lead to any confusion.  

\eject

\tableofcontents

\eject

\section*{LECTURE 1}
\addtocounter{section}{1}
\addcontentsline{toc}{section}{Lecture 1}

This Lecture is devoted to a  review of the relevant mathematical concepts, such as
 Lie algebroid, Courant bracket, Dirac structure and generalized complex geometry (also 
  its real analog).
 The presentation is rather sketchy and we leave many technical details aside. 

  For further reading on the Lie algebroids we recommend  \cite{Mak} and \cite{CW}. 
  On details of generalized complex geometry the reader may consult \cite{gualtieri}.

\subsection{Lie algebroid}
\label{subalgebroid}

Any course on the differential geometry starts from the introduction of $TM$, the tangent 
 bundle of smooth manifold $M$.  The sections of $TM$ are the vector fields. One of the 
  most important properties of $TM$ is that there exists a natural Lie bracket $\{\,\,,\,\,\}$
    between  vector fields.  The existence of a Lie bracket between vectors fields allows 
    the introduction of many interesting geometrical structures. Let us consider the example 
    of the complex structure:
    \begin{Exa}\label{exacomplex}
 An almost  complex structure $J$ on $M$ can be defined as
  a linear map (endomorphism)  $J: TM \rightarrow TM$ such that $J^2 = -1$. This allows us 
   to introduce the projectors
   $$ \pi_\pm= \frac{1}{2} ( 1\pm i J),\,\,\,\,\,\,\,\,\,\,\,\pi_+ + \pi_- =1,$$
   which induce a decomposition of complexified tangent space
  $$TM \otimes \mathbb{C} =  T^{1,0}M \oplus T^{0,1}M$$
   into a holomorphic and an antiholomorphic part, 
    $\pi_- v = v$ $v \in T^{(1,0)}M$ and $\pi_+ w = w$ $w \in T^{(0,1)}M$.
   The almost complex structure $J$ is integrable 
    if the subbundles $T^{(1,0)}M$ and $T^{(0,1)}M$  are involutive with respect to the Lie bracket, i.e. if
   $$ \pi_-\{\pi_+ v,  \pi_+ w\} =0,\,\,\,\,\,\,\,\,\,\,\,\,\,\,\,\,\,\,\,
   \pi_+\{\pi_- v,  \pi_- w\} =0$$ 
   for any $v, w \in \Gamma(TM)$. The manifold $M$ with such an integrable $J$ is called a complex 
    manifold. 
  \end{Exa}
 From this example we see that the Lie bracket plays a crucial role in the definition of  integrability 
  of a complex structure $J$.

   $TM$ is vector bundle with a Lie bracket. One can try to define a generalization of  $TM$
    as a vector bundle with a Lie bracket.  Thus we come now to the definition of a Lie algebroid 
\begin{Def}\label{defalgebroid}
A Lie algebroid is a vector bundle ${\mathbb L}$ over a manifold $M$ together with a
 bundle map (the anchor) $\rho: {\mathbb L} \rightarrow TM$ and a Lie bracket $\{\,\,,\,\,\}$ on 
  the space $\Gamma({\mathbb L})$ of sections of ${\mathbb L}$ satisfying
  $$ \rho \left ( \{ v, k\} \right ) = \{ \rho(v), \rho(k)\},\,\,\,\,\,\,\,\,\,\,\,\,\,v, k \in \Gamma({\mathbb L})$$
  $$ \{v, f k \} = f \{v,  k \} + (\rho(v) f) k,\,\,\,\,\,\,\,\,\,\,\,\,\,\,\,\,\,\,
   v, k \in \Gamma({\mathbb L}),\,\,\,f\in C^\infty (M)$$
\end{Def}
 In this definition $\rho(v)$ is a vector field and $(\rho(v) f)$ is the action of the vector field on 
  the function $f$, i.e. the Lie derivative of $f$ along $\rho(v)$.
 Thus the set of sections $\Gamma({\mathbb L})$ is a Lie algebra and there exists a Lie algebra homomorphism 
  from $\Gamma({\mathbb L})$ to $\Gamma(TM)$.

 To illustrate the definition \ref{defalgebroid}
we consider the following examples
\begin{Exa}\label{algtrivial}
 The tangent bundle $TM$ is a Lie algebroid with $\rho = id$.
 \end{Exa}
  \begin{Exa}\label{algsubbun}
  Any integrable subbundle ${\mathbb L}$ of $TM$ is Lie algebroid. The anchor map is inclusion 
  $$ {\mathbb L} \hookrightarrow TM$$
   and the Lie bracket on $\Gamma({\mathbb L})$ is given by the restriction of the ordinary Lie bracket 
    to ${\mathbb L}$. 
  \end{Exa}
    The notion of a Lie algebroid can obviously  be complexified.  
     For a complex Lie algebroid ${\mathbb L}$ we can use the same 
     definition \ref{defalgebroid} but with  ${\mathbb L}$ being a complex vector bundle and 
       the anchor map $\rho : {\mathbb L} \rightarrow TM \otimes {\mathbb C}$.
       \begin{Exa}\label{algcomplex}
       In example \ref{exacomplex} for the complex manifold $M$, $T^{(1,0)}M$ is an example 
        of a complex Lie algebroid with the anchor given by inclusion 
        $$ T^{(1,0)}M \hookrightarrow TM \otimes \mathbb{C}.$$
       \end{Exa} 
   
   It is instructive to rewrite the definition of Lie algebroid in local coordinates.  On a trivializing 
    chart we can choose the local coordinates $X^\mu$ ($\mu =1,...,\dim M$) and a basis $e^A$
     ($A=1,..., \rank \,{\mathbb L}$) on the fiber. In these local coordinates we introduce the anchor $\rho^{\mu A}$
     and the structure constants according to
     $$ \rho(e^A)(X) = \rho^{\mu A}(X) \d_\mu,\,\,\,\,\,\,\,\,\,\,\,\,\,\,\,\,\,
      \{ e^A, e^B\} = f^{AB}_{\,\,\,\,\,\,C} e^C.$$
    The compatibility conditions from the definition \ref{defalgebroid} imply the following 
     equation
     $$ \rho^{\nu A} \d_\nu \rho^{\mu B} - \rho^{\nu B} \d_\nu \rho^{\mu A} = f^{AB}_{\,\,\,\,\,\,C} \rho^{\mu C}$$
     $$\rho^{\mu [D}\d_\mu f^{AB]}_{\,\,\,\,\,\,C} + f^{[AB}_{\,\,\,\,\,\,L} f^{D]L}_{\,\,\,\,\,\,C} =0$$
    where $[\,\,]$ stands for the antisymmetrization.
   
   To any real Lie algebroid we can associate a characteristic foliation which is defined as follows.
   The image of anchor map $\rho$
   $$ \Delta = \rho ({\mathbb L}) \subset TM$$    
    is spanned by the smooth vector fields and thus it defines a smooth distribution.  Moreover this 
     distribution is involutive with the respect to the Lie bracket on $TM$. If the rank of this distribution 
     is constant then we can use the Frobenius theorem  and there exists a corresponding 
       foliation on $M$. However tha rank of $D$ does not have to be a constant and one should use
        the generalization of the Frobenius theorem due to Sussmann \cite{sussmann}.  
         Thus for any
   real Lie algebroid $\Delta_D = \rho ({\mathbb L})$ is integrable distribution in sense of Sussmann and there 
    exists a generalized foliation. 
   
  For  a complex Lie algebroid the situation is a bit more involved. The image of the anchor map
 $$\rho ({\mathbb L}) = E \subset TM \otimes {\mathbb C}$$
  defines two real distribution 
  $$ E + \bar{E} = \theta \otimes {\mathbb C},\,\,\,\,\,\,\,\,\,\,\,\,\,\,\,\,\,\,\,
   E \cap \bar{E} = \Delta \otimes {\mathbb C}.$$
   If $E + \bar{E} =TM \otimes {\mathbb C}$ then $\Delta$ is a smooth real distribution in the sense of Sussmann
    which defines a generalized  foliation.

\subsection{Geometry of $TM \oplus T^*M$}

At this point it would be natural to ask the following question. 
 How can one generate interesting examples of real and complex Lie
 algebroids? In this subsection we consider the tangent plus cotangent bundle 
  $TM \oplus T^*M$  or its complexification, $(TM\oplus T^*M)\otimes \mathbb{C}$ 
   and later we will show how one can construct  Lie algebroids 
  as subbundles of $TM \oplus T^*M$.   
 
 The section of  tangent plus cotangent bundle, $TM \oplus T^*M$, is a pair of objects, a vector 
  field $v$ and a one-form $\xi$. We adopt the following notation for a
  section: $v + \xi \in \Gamma (TM \oplus T^*M)$. 
     There exists a natural symmetric 
   pairing which is given by
   \beq
    \langle v + \xi, s+ \lambda \rangle = \frac{1}{2} (i_v \lambda + i_s \xi ),
   \eeq{pairingdefin}
    where $i_v \lambda$ is the contraction of a vector field $v$ with  one-form $\lambda$.
    In the local coordinates $(dx^\mu, \d_\mu)$ the pairing (\ref{pairingdefin}) can be rewritten
     in matrix form as
    \beq
   \langle A, B \rangle = \langle v + \xi, s+ \lambda \rangle = \frac{1}{2} \left ( \begin{array}{cc}
                                                                                                           v & \xi
                                                                                                           \end{array}\right )
                                                                                                           \left ( \begin{array}{cc}
                                                                                                              0 & 1\\
                                                                                                              1 & 0
                                                                                                        \end{array} \right ) 
                                                                                                        \left (\begin{array}{cc}
                                                                                                        s \\
                                                                                                        \lambda 
                                                                                                        \end{array} \right ) = A^t {\cal I} B,
                                                                                                          \eeq{pairinglocal}
  where
  $${\cal I} = \frac{1}{2}  \left ( \begin{array}{cc}
                                                                                                              0 & 1\\
                                                                                                              1 & 0
                                                                                                        \end{array} \right ) $$
 is a metric in a local coordinates  $(dx^\mu, \d_\mu)$.                                                                                                      
 ${\cal I}$ has   signature $(d,d)$ and thus here is natural action of $O(d,d)$ which preserves 
  the pairing.

The subbundle ${\mathbb L} \subset TM \oplus T^*M$ is called isotropic if $\langle A, B\rangle =0$ 
 for  all $A, B \in \Gamma({\mathbb L})$.  ${\mathbb L}$ is called maximally isotropic if 
 $$ \langle A, B\rangle =0,\,\,\,\,\,\,\,\,\,\,\,\,\,\,\,\,\,\,\,\forall A \in \Gamma({\mathbb L})$$
  implies that $B \in \Gamma({\mathbb L})$. 
 
 There is no  canonical  Lie bracket defined on the sections of $TM\oplus T^*M$. However 
  one can introduce  the following bracket
  \beq
   [ v + \xi, s + \lambda]_c = \{ v, s\} + {\cal L}_v \lambda - {\cal L}_s \xi -
   \frac{1}{2} d (i_v \lambda - i_s \xi ),
  \eeq{definiskakcour}
   which is called the Courant bracket. In (\ref{definiskakcour}) ${\cal L}_v$ stands for the Lie 
    derivative along $v$ and $d$ is de Rham differential on the forms. 
  The Courant bracket is antisymmetric and it does not satisfy the Jacobi identity. Nevertheless it is interesting 
    to examine how it fails to satisfy the Jacobi identity. Introducing the Jacobiator
    \beq
     Jac(A, B, C) = [ [A, B]_c, C]_c + [[B, C]_c, A]_c + [[C, A]_c B]_c
    \eeq{definjacobiator}
     one can prove the following proposition
    \begin{pro}\label{propjac}
  $$  Jac (A, B, C) = d(Nij(A, B, C))$$
     where 
     $$ Nij (A, B, C) = \frac{1}{3} \left ( \langle [A, B]_c, C\rangle + \langle [B, C]_c, A\rangle 
      + \langle [C, A]_c, B\rangle \right )$$
    and   where $A, B, C \in \Gamma(TM \oplus T^*M)$.
    \end{pro}
     {\bf Proof}: Let us sketch the main steps of the proof. 
     We  define the Dorfman bracket
     $$ (v +\omega) * (s + \lambda) = \{ v,s\} + {\cal L}_v \lambda - i_s d \omega ,$$
      such that its antisymmetrization 
      $$[ A, B]_c= A*B - B*A$$
       produces the Courant bracket. From the definitions of the Courant and Dorfman 
        brackets we can also deduce the following relation
       $$ [A, B]_c = A * B - d\langle A, B\rangle .$$
       It is crucial that the Dorfman bracket satisfies a kind of Leibniz rule
       $$ A * (B*C)= (A*B)*C + B * (A * C),$$
        which can be derived directly from the definition of the Dorfman bracket. 
         The combination of two last expressions leads to the formula 
         for the Jacobiator in the proposition. $\Box$

   Next we would like to investigate  the symmetries of the Courant bracket. 
    Recall that the symmetries of the Lie bracket on $TM$ are described in 
     terms of bundle automorphism
$$\begin{array}{ccc}
  TM & \stackrel{F}{\longrightarrow} & TM\\
\Big{\downarrow}
 && \Big\downarrow\\
  M & \stackrel{f}{\longrightarrow} &M
\end{array}$$
 such that
$$ F(\{ v, k\} )= \{ F(v), F(k)\}.$$
 For the Lie bracket on $TM$ the only symmetry  is diffeomorphism, i.e. $F= f_*$. 
 
 Analogously we look for the symmetries of the Courant bracket as bundle 
  automorphism
  $$\begin{array}{ccc}
  TM \oplus T^*M& \stackrel{F}{\longrightarrow} & TM\oplus T^*M\\
\Big{\downarrow}
 && \Big\downarrow\\
  M & \stackrel{f}{\longrightarrow} &M
\end{array}$$
such that 
$$ [F(A), F(B)]_c = F([A, B]_c),\,\,\,\,\,\,\,\,\,\,\,\,\,\,\,A,B \in \Gamma(TM\oplus T^*M)$$
 and in addition we require that it preserves the natural pairing $\langle\,\,,\,\,\rangle$.
Obviously  $Diff(M)$ is the symmetry of the Courant bracket with $F=f_* \oplus f^*$.
 However there exists an additional symmetry.  For any two-form $b \in \Omega^2(M)$
  we can define the transformation
  \beq
 e^b (v + \lambda)  \equiv  v + \lambda + i_v b ,
 \eeq{actionofB}
 which  preserves the pairing. Under this transformation the Courant bracket transforms as 
  follows 
 \beq
  [ e^b (v + \xi), e^b(s + \lambda) ]_c = e^b ([v+ \xi, s+\lambda]) + i_v i_s db .
 \eeq{btarsnfkrksoo99}
 If $db=0$ then we have a an orthogonal symmetry of the Courant bracket.  Thus we arrive to 
  the following proposition \cite{gualtieri} :
  \begin{pro}\label{symcourant}
  The group of orthogonal Courant automorphisms of $TM \oplus T^*M$ is semi-direct product 
   of $Diff(M)$ and $\Omega_{closed}^2(M)$.  
 \end{pro}

$TM\oplus T^*M$ equipped with the natural pairing $\langle\,\,,\,\,\rangle$ and the Courant bracket $[\,\,,\,\,]_c$ is 
 an example of the Courant algebroid.  In general the Courant algebroid is a vector bundle with 
  the bracket $[\,\,,\,\,]_c$ and the pairing $\langle \,\,,\,\,\rangle$ which satisfy the same 
    properties we have described in this subsection.

\subsection{Dirac structures}

   In this subsection we will use the properties of $TM\oplus T^*M$ in order to construct 
    the examples of real and complex Lie algebroids.
  
   The proposition \ref{propjac} implies the following immediate corollary 
  \begin{cor}\label{cordirac}
 For maximally isotropic subbundle  ${\mathbb L}$ of  $TM \oplus T^*M$ or $(TM \oplus T^*M)\otimes \mathbb{C}$
  the following three statements are equivalent
  \begin{itemize}
 \item[*] ${\mathbb L}$ is involutive 
 \item[*]  $Nij|_{{\mathbb L}} =0$
\item[*]   $Jac|_{{\mathbb L}}=0$
   \end{itemize}
  \end{cor}
 Here  we call ${\mathbb L}$  involutive if for any $A,B \in \Gamma({\mathbb L})$  the bracket 
  $[A, B]_c \in \Gamma({\mathbb L})$. 
  \begin{Def}\label{defdirac}
 An involutive maximally isotropic subbundle ${\mathbb L}$ of $TM \oplus T^*M$ 
   (or $(TM \oplus T^*M)\otimes \mathbb{C}$) is called a real (complex) Dirac structure. 
   \end{Def}
 It follows from corollary  \ref{cordirac} that ${\mathbb L}$ is a Lie algebroid with the bracket given 
  by the restriction of the Courant bracket to ${\mathbb L}$. Since $Jac|_{\mathbb L}=0$ the bracket $[\,\,,\,\,]_c|_{{\mathbb L}}$ is 
   a Lie bracket. The anchor map is given by a natural projection to $TM$.

 Let us consider some examples of Dirac structures
 
 \begin{Exa}
 The tangent bundle $TM \subset TM \oplus T^*M$ is a Dirac structure since $TM$ is a maximally 
  isotropic subbundle. Moreover  the restriction 
  of the Courant bracket to $TM$ is the standard Lie bracket on $TM$ and thus it is an involutive 
   subbundle.
 \end{Exa}
 
 \begin{Exa}
  Take a two-form $\omega \in \Omega^2(M)$ and consider the following subbundle of $TM\oplus T^*M$
 $${\mathbb L}= e^{\omega} (TM) = \{ v + i_v \omega, v\in TM\} .$$
  This subbundle is maximally isotropic since $\omega$ is a two-form. Moreover one can show that 
 ${\mathbb L}$ is involutive if   $d\omega=0$. Thus if  $\omega$ is a presymplectic structure\footnote{The two-form $\omega$ 
  is called a symplectic   structure if $d\omega=0$ and $\exists\,\, \omega^{-1}$. If two-form is just closed
   then it is called a presymplectic structure.} then ${\mathbb L}$ is an example
  of a real Dirac structure. 
 \end{Exa}
 
 \begin{Exa}
  Instead we can take an antisymmetric bivector $\beta \in \Gamma (\wedge^2 TM)$ and define 
   the subbundle 
$$  {\mathbb L}= \{ i_\beta \lambda  + \lambda, \lambda\in T^*M\} ,$$
 where $i_\beta \lambda$ is a contraction of bivector $\beta$ with one-form $\lambda$. 
${\mathbb L}$ is involutive when $\beta$ is a Poisson structure\footnote{The antisymmetric bivector $\beta^{\mu\nu}$ is called Poisson if it satisfies $\beta^{\mu\nu}\d_\nu \beta^{\rho\sigma} + \beta^{\rho\nu} \d_\nu \beta^{\sigma\mu}
 + \beta^{\sigma\nu} \d_\nu \beta^{\mu\rho}=0$. The name of $\beta$ is justified by the fact that 
 $\{ f, g\} =(\d_\mu f) \beta^{\mu\nu} (\d_\nu g)$ defines a Poisson bracket for $f,g \in C^\infty(M)$.}. Thus for a Poisson manifold ${\mathbb L}$ is a real 
 Dirac structure. 
 \end{Exa}
 
 \begin{Exa}\label{ecoma11}
  Let $M$ to be a complex manifold and consider the following subbundle 
   of $(TM \oplus T^*M)\otimes \mathbb{C}$
  $$ {\mathbb L} = T^{(0,1)}M \oplus T^{*(1,0)}M$$
   with the sections being antiholomorphic vector fields plus holomorphic forms. ${\mathbb L}$ is maximally isotropic
   and involutive (this follows immediately when $[\,\,,\,\,]_c|_{{\mathbb L}}$ is written explicitly). Thus for a complex
    manifold,  ${\mathbb L}$ is an example of a complex Dirac structure. 
 \end{Exa}

\subsection{Generalized complex structures}

 In this subsection we present the central notion for us, a generalized complex structure. 
  We will present the different but equivalent definitions and discuss some basic examples
   of a generalized complex structure.

 We have defined all basic notions needed for the definition of a generalized complex structure
\begin{Def}\label{defGCS}
 The generalized complex structure is a complex Dirac structure ${\mathbb L} \subset (TM \oplus T^*M) \otimes \mathbb{C}$ such that ${\mathbb L} \cap \bar{\mathbb L} =\{0\}$ .
\end{Def}
 In other words a generalized complex structure gives us a decomposition 
 $$ (TM \oplus T^*M) \otimes \mathbb{C} = {\mathbb L} \oplus \bar{\mathbb L}$$
  where ${\mathbb L}$ and $\bar{\mathbb L}$ are complex Dirac structures. 
 
  There exist an alternative definition however. Namely 
  we can mimic the standard description of the usual complex structure which can be  
   defined as an endomorphism $J: TM \rightarrow TM$ with additional properties,
  see Example \ref{exacomplex}. 
    
  Thus  in analogy we define the endomorphism  
 $$ {\cal J}: TM \oplus T^*M \rightarrow TM \oplus T^*M,$$
  such that
   \beq
  {\cal J}^2 = - 1_{2d} .
  \eeq{definall282929}
    There exist  projectors
    $$ \Pi_\pm = \frac{1}{2} \left ( 1_{2d} \pm i {\cal J} \right )$$
     such that $\Pi_+$ is projector for $\bar{\mathbb L}$ and $\Pi_-$ is the projector for ${\mathbb L}$.  However
      ${\mathbb L}$ ($\bar{\mathbb L}$) is a maximally isotropic subbundle of $(TM \oplus T^*M) \otimes \mathbb{C}$.
       Thus we need to impose a compatibility condition between the natural pairing and 
        ${\cal J}$ in order to insure that ${\mathbb L}$ and $\bar{\mathbb L}$ are maximally isotropic spaces.
         Isotropy of ${\mathbb L}$ implies that for any sections
           $A, B \in \Gamma((T \oplus T^*)\otimes\mathbb{C})$
   $$ \langle\Pi_-  A , \Pi_- B \rangle = A^t \Pi_-^t {\cal I} \Pi_- B = \frac{1}{4}
    A^t ( {\cal I} + i {\cal J}^t {\cal I} + i {\cal I}{\cal J} - {\cal J}^t {\cal I} {\cal J}) B=0 $$
     which produces the following condition 
   \beq
    {\cal J}^t {\cal I} = - {\cal I} {\cal J} .
   \eeq{djksdkkas000-}
    If there exists a ${\cal J}$ satisfying (\ref{definall282929}) and (\ref{djksdkkas000-})
     then we refer to ${\cal J}$ as an almost generalized complex structure. Next we have 
      to add the integrability conditions, namely that ${\mathbb L}$ and $\bar{\mathbb L}$ are involutive 
   with respect to the Courant bracket, i.e.
   \beq
    \Pi_{\mp} [ \Pi_\pm A, \Pi_\pm B ]_c =0
   \eeq{inegrabilityweoop29}
    for any sections $A, B \in \Gamma (TM \oplus T^*M)$.
   Thus  ${\mathbb L}$ is $+i$-egeinbundle of ${\cal J}$ and $\bar{\mathbb L}$ is $-i$-egeinbundle of ${\cal J}$. 
    To summarize a generalized complex structure can be defined as an endomorphism ${\cal J}$
     with the properties  (\ref{definall282929}), (\ref{djksdkkas000-}) and (\ref{inegrabilityweoop29}).
 
   An endomorphism ${\cal J} : TM \oplus T^*M \rightarrow TM\oplus T^*M$ satisfying
    (\ref{djksdkkas000-}) can be written in the form
   \beq
   {\cal J} = \left ( \begin{array}{cc}
      J & P \\
      L & - J^t 
      \end{array} \right )
   \eeq{formgenakksl}
    with $J: TM \rightarrow TM$, $P: T^*M \rightarrow TM$, $L: TM \rightarrow T^*M$ and 
     $J^t: T^*M \rightarrow TM$. Indeed $J$ can be identified  with a $(1,1)$-tensor, $L$ with a two-form
      and $P$ with an antisymmetric bivector.  Imposing further the conditions (\ref{definall282929}) 
       and (\ref{inegrabilityweoop29}) we arrive to the set of algebraic and differential conditions on 
        the tensors $J$, $L$ and $P$ which were first studied in  \cite{Lindstrom:2004iw}. 

To illustrate the definition of a generalized complex structure we consider a few examples.
 
\begin{Exa}\label{exa123}
 Consider ${\cal J}$  of the following form
 $$ {\cal J} = \left ( \begin{array}{cc}
      J & 0 \\
      0 & - J^t 
      \end{array} \right ).$$
   Such   ${\cal J}$ is a generalized complex structure if and only if $J$ is a complex structure.
 The corresponding Dirac structure is 
    $$ {\mathbb L}= T^{(0,1)}M \oplus T^{*(1,0)}M$$
 as in example \ref{ecoma11}.
\end{Exa}

\begin{Exa}\label{exa1234}
 Consider a ${\cal J}$ of the form
 $$ {\cal J} = \left ( \begin{array}{cc}
      0 & -\omega^{-1} \\
      \omega &  0
      \end{array} \right ).$$
       Such ${\cal J}$ is a generalized complex structure if and only if 
      $\omega$ is a symplectic structure. The corresponding Dirac structure is 
       defined as follows
      $$ {\mathbb L} =\{ v -i ( i_v \omega) ,\,v \in TM \otimes \mathbb{C}  \}. $$
\end{Exa}

\begin{Exa}
 Consider a generic generalized complex structure ${\cal J}$ written in the form (\ref{formgenakksl}). 
  Investigation of  the conditions  (\ref{definall282929}) 
       and (\ref{inegrabilityweoop29}) leads to the fact that  
 $P$ is a Poisson tensor. Furthermore one can show that locally there is a  symplectic foliation
   with a transverse complex structure. Thus locally a generalized complex manifold is a product 
    a symplectic and complex manifolds \cite{gualtieri}. The dimension of the generalized complex manifold is 
     even. 
\end{Exa}

\subsection{Generalized product structure}

 Both complex structure and generalized complex structures have real analogs.  In this subsection
  we will discuss them briefly. Some of the observations presented in this subsection are original. However 
   they follow rather straightforwardly from a slight modification of the complex case.  
  
   The complex structure described in the example \ref{exacomplex} has a real analog which is called 
   a product structure \cite{yano}
    
   \begin{Exa}\label{exaproduct}
 An almost  product structure $\Pi$ on $M$ can be defined as
  a map $\Pi: TM \rightarrow TM$ such that $\Pi^2 = 1$. This allows us 
   to introduce the projectors
   $$ \pi_\pm= \frac{1}{2} ( 1\pm \Pi),\,\,\,\,\,\,\,\,\,\,\,\pi_+ + \pi_- =1,$$
   which induce the decomposition of real tangent space
  $$TM  =  T^+M \oplus T^-M$$
   into two parts, 
    $\pi_- v = v$ $v \in T^+M$ and $\pi_+ w = w$ $w \in T^-M$.
     The dimension of $T^+M$ can be different from the dimension of $T^-M$ and thus
      the manifold $M$ does not have to be even dimensional. 
   The almost product structure $\Pi$ is integrable 
    if the subbundles $T^{+}M$ and $T^{-}M$  are involutive with respect to the Lie bracket, i.e.
   $$ \pi_-\{\pi_+ v,  \pi_+ w\} =0,\,\,\,\,\,\,\,\,\,\,\,\,\,\,\,\,\,\,\,
   \pi_+\{\pi_- v,  \pi_- w\} =0$$ 
   for any $v, w \in \Gamma(TM)$.  We refer to an integrable almost product structure as
    product structure.
   A manifold $M$ with such integrable $\Pi$ is called a locally product 
    manifold. 
  \end{Exa}   
  There exists always the trivial example of such structure $\Pi= id$.

 Obviously the definition \ref{defGCS} of generalized
 complex structure also has a real analog.
 \begin{Def}\label{defgenproduct}
 A generalized product structure is a pair  of real Dirac structures ${\mathbb L}_\pm$ 
  such that ${\mathbb L}_+ \cap {\mathbb L}_- =\{0\}$. In other words
$$ TM \oplus T^*M= {\mathbb L}_+ \oplus {\mathbb L}_-$$
\end{Def}
 Indeed the definitions \ref{defGCS} and \ref{defgenproduct} are examples of complex and real Lie 
   bialgebroids \cite{bialgebroid}.  However we will not discuss this structure here.

 Analogously to the complex case we can define an almost generalized product structure by means 
  of an endomorphims
  $$ {\cal R} :  TM \oplus T^*M \rightarrow TM\oplus T^*M$$
   such that 
   \beq
   {\cal R}^2 = 1_{2d},
   \eeq{djkslsllooo}
   and 
   \beq
   {\cal R}^t {\cal I} = - {\cal I} {\cal R}.
   \eeq{bdlallllll}
 The corresponding projectors
 $$ p_\pm = \frac{1}{2} (1_{2d} \pm {\cal R})$$
 define two maximally isotropic subspaces ${\mathbb L}_+$ and ${\mathbb L}_-$. The integrability conditions
 are  given by
  \beq
   p_\mp [ p_\pm A, p_\pm B]_c = 0 ,
  \eeq{dwkkk29937}
   where $A$ and $B$ are any sections of $TM\oplus T^*M$.  In analogy with (\ref{formgenakksl})
    we can write an endormorphism which satisfies (\ref{bdlallllll}) as follows
     \beq
   {\cal R} = \left ( \begin{array}{cc}
      \Pi &  \tilde{P} \\
      \tilde{L}  & - \Pi^t 
      \end{array} \right ),
   \eeq{formgenakkslextra}
    where $\Pi$ is a $(1,1)$-tensor, $\tilde{P}$ is an antisymmetric bivector and $\tilde{L}$ is a two-form.
     The conditions (\ref{djkslsllooo}) and (\ref{dwkkk29937})  imply  similar algebraic and the same 
      differential conditions for the tensors $\Pi$, $\tilde{L}$ and $\tilde{P}$ as in \cite{Lindstrom:2004iw}.
    
   Let us give a few examples of a generalized product structure.
   \begin{Exa}
    Consider ${\cal R}$  of the following form
 $$ {\cal R} = \left ( \begin{array}{cc}
      \Pi & 0 \\
      0 & - \Pi^t 
      \end{array} \right ).$$
   Such   an ${\cal R}$ is a generalized product structure if and only if $\Pi$ is a standard product structure.
    This example justifies the name, we have proposed: a generalized product structure.
     The Dirac structure ${\mathbb L}_+$ is
     $$ {\mathbb L}_+ = T^{+}M \oplus T^{*-}M,$$
      where $\lambda \in T^{*-}M$ if $\pi_+ \lambda =\lambda$, see Example \ref{exaproduct}.
\end{Exa}
\begin{Exa}
 Consider an ${\cal R}$ of the form
 $$ {\cal R} = \left ( \begin{array}{cc}
      0 &  \omega^{-1} \\
      \omega &  0
      \end{array} \right ).$$
       Such an ${\cal R}$ is  a generalized product structure if and only if 
      $\omega$ is a symplectic structure.
\end{Exa}

 For the generic generalized product structure ${\cal R}$ (\ref{formgenakkslextra})
      $\tilde{P}$ is a Poisson structure. Generalizing the complex case one 
      can show that locally there is a symplectic foliation with a transverse product 
       structure. 
 Thus  locally  a generalized product manifold is a product of symplectic and locally product
  manifolds.

\subsection{Twisted case}

 Indeed one can construct on $TM \oplus T^*M$ more than one bracket with the same properties as 
 the Courant bracket.  Namely the different brackets are parametrized by a closed three form
 $H\in \Omega^3(M)$, $dH=0$ and are defined as follows
  \beq
  [v + \xi, s + \lambda]_H = [v + \xi, s +\lambda ]_c + i_v i_s H.
 \eeq{defintwistedka}
  We refer to this bracket as the twisted Courant bracket.  This bracket has the same properties
   as the Courant bracket.  If $H=db$ then the last term on the right hand side of
     (\ref{defintwistedka}) can be generated by non-closed b-transform, see (\ref{btarsnfkrksoo99}). 
   
   Thus we can define a twisted Dirac structure, a twisted
   generalized complex structure and a twisted generalized product structure.   
    In all definitions the Courant bracket $[\,\,,\,\,]_c$ should be replaced by the twisted 
     Courant bracket $[\,\,,\,\,]_H$. For example, a twisted generalized complex structure
    ${\cal J}$ satisfies
     (\ref{definall282929})  and (\ref{djksdkkas000-}) and now the integrability is defined 
      with respect to twisted Courant bracket as
    \beq
    \Pi_{\mp} [ \Pi_\pm (v +\xi ), \Pi_\pm (s + \lambda) ]_H =0. 
   \eeq{inegrabilityweoop29new}
   
 There is a nice relation of the twisted version to gerbes \cite{gualtieri, hitchinlast}. However due to lack of time we will have to leave
  it aside. 
 
\section*{LECTURE 2}
\addtocounter{section}{1}
\addcontentsline{toc}{section}{Lecture 2}

In this Lecture we turn our attention to physics. In particular we would like to 
 show that the  mathematical notions introduced in Lecture 1 appear naturally 
  in the context of string theory. 
  Here  we focus on the classical aspect of the 
  hamiltonian formalism for the world-sheet  theory.  

\subsection{String phase space $T^*LM$}

 A wide class of sigma models share the following phase space description. 
 For the world-sheet $\Sigma = S^1 \times {\mathbb R}$ 
 the phase space can be identified with  a cotangent bundle $T^*LM$ of the loop space 
 $LM=\{ X: S^1 \rightarrow M\}$. Using local coordinates $X^\mu(\sigma)$ and their conjugate 
 momenta $p_\mu(\sigma)$ 
 the standard symplectic form on $T^*LM$ is given by
\beq
 \omega = \int\limits_{S^1} d\sigma \,\, \delta X^\mu \wedge \delta p_\mu ,
\eeq{sympbos}
 where $\delta$ is de Rham differential on $T^*LM$ and $\sigma$ is a coordinate along $S^1$.
 The symplectic form (\ref{sympbos}) can be twisted by a closed three form $H \in \Omega^3(M)$, $dH=0$
 as follows 
\beq
 \omega = \int\limits_{S^1} d\sigma \,\,(\delta X^\mu \wedge \delta p_\mu +  H_{\mu\nu\rho} \d X^\mu 
 \delta X^\nu \wedge \delta X^\rho ) ,
\eeq{symptwist}
 where $\d \equiv \d_\sigma$ is derivative with respect to $\sigma$. 
 For both symplectic structures the following transformation is canonical 
\beq
X^\mu\,\,\rightarrow\,\,X^\mu,\,\,\,\,\,\,\,\,\,\,\,\,\,\,\,\,\,\,\,
 p_\mu\,\,\rightarrow\,\,p_\mu + b_{\mu\nu} \d X^\nu
\eeq{canobtransf}
 associated with a closed two form, $b \in \Omega^2(M)$,  $db=0$. There are also canonical 
 transformations which correspond to $Diff(M)$ when $X$ transforms as a coordinate and 
 $p$ as a section of the cotangent bundle $T^*M$. In fact the group of local 
  canonical transformations\footnote{By local canonical transformation we mean
 those canonical transformations where the new pair $(\tilde{X},\tilde{p})$ is given as
  a local expression in terms of the old one $(X, p)$. For example, in the discussion of T-duality 
 one uses non-local canonical transformations, i.e. $\tilde{X}$ is a non-local expression in terms of $X$.}
 for $T^*LM$ is a semidirect product of $Diff(M)$ and $\Omega^2_{closed}(M)$.  
  Therefore we come to the following proposition
  \begin{pro}\label{symcourantextra}
  The group of local canonical transformations on $T^*LM$ is isomorphic to the 
 group of orthogonal  automorphisms of Courant bracket. 
 \end{pro}
 
 See the proposition \ref{symcourant} and  the discussion of 
  the symmetries on the Courant bracket in the previous Lecture. 
  The proposition \ref{symcourantextra} is a first 
  indication that the geometry of $T^*LM$ is related to the generalized geometry of $TM\oplus T^*M$.

\subsection{Courant bracket and $T^*LM$}
\label{cour}

Indeed the Courant bracket by itself can be "derived" from $T^*LM$. Here we present a nice 
 observation on the relation between the Courant bracket and the Poisson bracket on 
  $C^\infty(T^*LM)$  which is due to \cite{Alekseev:2004np}. 

Let us define for any section $(v+\xi) \in \Gamma(TM\oplus T^*M)$ (or its complexified version)
 a current (an element of $C^\infty(T^* LM)$) as follows
\beq
 J_\epsilon (v+ \xi ) = \int\limits_{S^1} d\sigma\,\,\epsilon (v^\mu p_\mu + \xi_\mu \d X^\mu), 
\eeq{defcueoep}
 where $\epsilon \in C^\infty (S^1)$ is a test function. Using the symplectic structure (\ref{sympbos})
 we can calculate the Poisson bracket between two currents
\beq
\{ J_{\epsilon_1}(A),  J_{\epsilon_2}(B) \} = - J_{\epsilon_1 \epsilon_2} 
 ([A,B]_c ) + \int\limits_{S^1} d\sigma\,\, (\epsilon_1 \d \epsilon_2 - \epsilon_2
 \d \epsilon_1) \langle A,B \rangle ,
\eeq{definslll2w92992}
 where $A, B \in \Gamma (TM \oplus T^*M)$. 
On the right hand side of (\ref{definslll2w92992})  the Courant bracket and natural pairing
 on $TM\oplus T^*M$ appear. It is important to stress that the Poisson bracket $\{\,\,,\,\,\}$
  is associative while the Courant bracket $[\,\,,\,\,]_c$ is not. 
 
  If we consider ${\mathbb L}$ to be a real (complex) Dirac structure (see definition \ref{defdirac}) then 
    for $A, B \in \Gamma({\mathbb L})$
    \beq
     \{ J_{\epsilon_1}(A), J_{\epsilon_2}(B) \} = - J_{\epsilon_1\epsilon_2} ([A, B]_c|_{\mathbb L}) ,
    \eeq{condk388390-1}
 where $[\,\,,\,\,]_c|_{\mathbb L}$ is the restriction of the Courant bracket to ${\mathbb L}$. Due to the isotropy of ${\mathbb L}$ the last term 
  on the right hand side of (\ref{definslll2w92992}) vanishes and $[\,\,, \,\,]_c|_{\mathbb L}$
    is a Lie bracket on $\Gamma({\mathbb L})$.  Thus there is a natural relation between the Dirac structures 
     and the current algebras.
 
 For any real (complex) Dirac structure ${\mathbb L}$ we can define the set of constraints in $T^*LM$
\beq
v^\mu p_\mu + \xi_\mu \d X^\mu = 0 ,
\eeq{firjsklllsl2929}
 where $(v+\xi) \in \Gamma({\mathbb L})$.  The conditions (\ref{firjsklllsl2929}) are first class constraints
   due to (\ref{condk388390-1}), i.e. they define a coisotropic submanifold of $T^*LM$.   
   Moreover the number of independent constraints is equal to $\dim {\mathbb L} = \dim M$ and 
    thus the constraints (\ref{firjsklllsl2929}) correspond to a topological field theory (TFT). 
     Since ${\mathbb L}$ is maximally isotropic  it then  follows from (\ref{firjsklllsl2929}) that 
   \beq
    \left ( \begin{array}{c}
     \d X\\
     p
     \end{array} \right ) \,\,\in\,\,X^*({\mathbb L}),
   \eeq{defindlla78830}
 i.e. $\d X + p$ take values in  the subbundle ${\mathbb L}$ (more precisely, in the pullback of ${\mathbb L}$). 
  The set (\ref{defindlla78830}) is equivalent to (\ref{firjsklllsl2929}). 
 Thus  with any real (complex) Dirac structure we can associate a classical TFT. 
     
 Also we could calculate the bracket (\ref{definslll2w92992}) between the currents using the symplectic 
  structure (\ref{symptwist}) with $H$. In this case the Courant bracket should be replaced by the 
   twisted Courant bracket.  Moreover we have to consider the twisted Dirac structure instead 
    of a Dirac structure. Otherwise all statement will remain true.

\subsection{String super phase space $T^*{\cal L}M$}

Next we would like to extend our construction and add odd partners to the fields $(X,p)$. 
 This will allow us to introduce more structure. 
 
 Let $S^{1,1}$ be a "supercircle" with coordinates $(\sigma, \theta)$, where $\sigma$ is a coordinate 
  along $S^1$ and $\theta$ is odd parter of $\sigma$ such that $\theta^2 =0$. 
 Then the corresponding 
 superloop space  is the space of maps, ${\cal L}M = \{ \Phi :  S^{1,1} \rightarrow M\}$. The phase space 
 is given by the cotangent bundle $\Pi T^*{\cal L}M$ of ${\cal L}M$, however with 
  reversed parity on the fibers. In what follows we use the letter "$\Pi$" to describe 
   the reversed parity on the fibers.
  Equivalently we can describe the space $\Pi T^* {\cal L}M$ as the space of maps
   $$ \Pi T S^1 \rightarrow \Pi T^*M, $$ 
 where  the supermanifold $\Pi TS^1$ ($\equiv S^{1,1}$)  is the tangent bundle of $S^1$ with reversed 
  parity of the fiber and the supermanifold $\Pi T^* M$ is the cotangent bundle of $M$
   with reversed parity on the fiber. 
   
   In local coordinates we have a scalar superfield $\Phi^\mu(\sigma,\theta)$
 and a conjugate momentum, spinorial superfield  $S_\mu(\sigma,\theta)$ with the following
 expansion
\beq
 \Phi^\mu (\sigma, \theta) = X^\mu(\sigma) + \theta \lambda^\mu(\sigma),\,\,\,\,\,\,\,\,\,\,\,\,\,\,\,\,\,\,\,\,\,\,
 S_\mu (\sigma, \theta)= \rho_\mu(\sigma) +i \theta p_\mu (\sigma),
\eeq{superfiel}
 where $\lambda$ and $\rho$ are fermions.  $S$ is a section of the pullback $X^*(\Pi T^*M)$
  of the cotangent bundle of $M$, considered as an odd bundle.
 The corresponding symplectic structure
 on $\Pi T^*{\cal L}M$ is 
\beq
 \omega = i \int\limits_{S^{1,1}} d\sigma d\theta\,\,(\delta S_\mu \wedge \delta\Phi^\mu -
  H_{\mu\nu\rho} D\Phi^\mu \delta \Phi^\nu \wedge \delta \Phi^\rho ),
\eeq{sympsuper}
 such that  after integration over $\theta$ the bosonic part of (\ref{sympsuper}) coincides with (\ref{symptwist}). 
    
    The above symplectic structure makes $C^{\infty}(\Pi T^*{\cal L}M)$ (the space of smooth 
   functionals on $\Pi T^*{\cal L}M$) into superPoisson algebra.  The space $C^{\infty}(\Pi T^*{\cal L}M)$
    has a natural $\mathbb{Z}_2$ grading with $|F|=0$ for even and $|F|=1$ for odd functionals. 
  For a functional $F(S, \phi)$ we define the left and right functional derivatives 
  as follows
 \beq
 \delta F = \int d\sigma d\theta \left ( \frac{ F \overleftarrow{\delta}}{\delta S_\mu} \delta S^\mu +
  \frac{F \overleftarrow{\delta}}{\delta \phi^\mu} \delta \phi^\mu \right ) =
  \int d\sigma d\theta \left ( \delta S_\mu \frac{ \overrightarrow{\delta} F}{\delta S_\mu}  +
  \delta \phi^\mu \frac{\overrightarrow{\delta} F}{\delta \phi^\mu}  \right ) .
 \eeq{Pdefinals;a[[[}
  Using this definition the Poisson bracket corresponding to (\ref{sympsuper})  with $H=0$ is given by
  \beq
  \{ F, G\} = i \int d\sigma d\theta \left ( \frac{ F \overleftarrow{\delta}}{\delta S_\mu} \frac{\overrightarrow{\delta} G}{\delta \phi^\mu} - \frac{F \overleftarrow{\delta}}{\delta \phi^\mu} \frac{\overrightarrow{\delta} G}
  {\delta S_\mu} \right ). 
  \eeq{definape8349}
  and with $H \neq 0$
     \beq
  \{ F, G\}_H = i \int d\sigma d\theta \left ( \frac{ F \overleftarrow{\delta}}{\delta S_\mu} \frac{\overrightarrow{\delta} G}{\delta \phi^\mu} - \frac{F \overleftarrow{\delta}}{\delta \phi^\mu} \frac{\overrightarrow{\delta} G}
  {\delta S_\mu}  + 2  \frac{ F \overleftarrow{\delta}}{\delta  S_\nu} H_{\mu\nu\rho} D\phi^\mu 
   \frac{\overrightarrow{\delta} G}{\delta S_\rho} \right  ). 
  \eeq{definape8349moreysj}
   These brackets $\{\,\,,\,\,\}$ and $\{\,\,,\,\,\}_H$
    satisfy the appropriate graded versions of antisymmetry, of the Leibnitz rule and 
    of the Jacobi identity
   \beq
    \{ F, G\} = - (-1)^{|F| |G|} \{G, F\},
  \eeq{ansiruwiqw99}
  \beq
   \{ F, GH \} = \{ F, G\} H + (-1)^{|F| |G|} G \{ F, H\},
  \eeq{Leibniziao2290}
  \beq
  (-1)^{|H||F|} \{ F, \{G, H\}\} + (-1)^{|F| |G|} \{ G, \{ H, F\}\} + (-1)^{|G| |H|} \{ H, \{ F, G \} \} =0.
  \eeq{Javoao20021}

 Next on $\Pi T S^1$  we have two natural operations, $D$ and $Q$.
   The derivative $D$ is defined as
   \beq
     D= \frac{\d }{\d \theta} + i \theta \d
   \eeq{definsoldfkfk}
    and the operator $Q$ as
    \beq
     Q = \frac{\d}{\d \theta} - i \theta \d.
    \eeq{djkfdhe892902}
    $D$ and $Q$ satisfy the following algebra
    \beq
     D^2 = i\d,\,\,\,\,\,\,\,\,\,\,\,\,\,\,\,\,\,
      Q^2 = - i\d,\,\,\,\,\,\,\,\,\,\,\,\,\,\,\,\,\,
      DQ + QD = 0 .
    \eeq{alshjwopqp[p}
     Here $\d$ stands for the derivative along the loop, i.e. along $\sigma$. 

 Again as in the purely bosonic case (see the proposition \ref{symcourantextra}) the group
   of local canonical transformations of $\Pi T^*{\cal L}M$ is 
 a semidirect product of $Diff(M)$ and $\Omega^2_{closed}(M)$.
 The $b$-transform now is given by
\beq
 \Phi^\mu\,\,\rightarrow\,\,\Phi^\mu,\,\,\,\,\,\,\,\,\,\,\,\,\,\,\,\,\,\,\,
 S_\mu\,\,\rightarrow\,\,S_\mu - b_{\mu\nu} D\Phi^\nu,
\eeq{canonfr;as}
with $b \in \Omega_{closed}^2(M)$. Moreover the discussion from subsection \ref{cour}
 can be generalized to the supercase. 

Consider first $C^\infty (\Pi T^*{\cal L}M)$ with $\{\,\,,\,\,\}$.  By construction 
 of $\Pi T^*{\cal L}M$ there exists the following generator
\beq
 {\mathbf Q}_1(\epsilon) = - \int\limits_{S^{1,1}} d\sigma d\theta\,\, \epsilon S_\mu Q \Phi^\mu ,
\eeq{manisfetsssu}
 where $Q$ is the operator introduced in (\ref{djkfdhe892902}) 
  and $\epsilon$ is an odd parameter (odd test function). 
 Using (\ref{sympsuper}) we can calculate the Poisson brackets for these generators
\beq
 \{ {\mathbf Q}_1(\epsilon), {\mathbf Q}_1(\tilde{\epsilon})\} = {\mathbf P}(2\epsilon\tilde{\epsilon}),
\eeq{susyalg1}
 where $P$ is the generator of translations along $\sigma$
\beq
 {\mathbf P}(a) = \int\limits_{S^{1,1}} d\sigma d\theta\,\, a S_\mu \d \Phi^\mu
\eeq{defintranms}
 with $a$ being an even parameter.  In physics such a generator  ${\mathbf Q}_1(\epsilon)$ 
  is called a supersymmetry generator and it has the meaning of a square root of the translations, see
   (\ref{susyalg1}).  Furthermore we call it a manifest supersymmetry since it exits as part
    of the superspace formalism.  One can construct a similar generator of manifest supersymmetry 
     on $C^\infty (\Pi T^*{\cal L}M)$ with $\{\,\,,\,\,\}_H$.  

\subsection{Extended supersymmetry and generalized complex structure}

 Consider $C^\infty (\Pi T^*{\cal L}M)$ with $\{\,\,,\,\,\}$.
 We look for a second supersymmetry generator. The second supersymmetry should 
 be generated by some ${\mathbf Q}_2(\epsilon)$ such that it satisfies the following brackets
\beq
 \{ {\mathbf Q}_1(\epsilon), {\mathbf Q}_2(\tilde{\epsilon})\} = 0,\,\,\,\,\,\,\,\,\,\,\,\,\,\,
 \{ {\mathbf Q}_2(\epsilon), {\mathbf Q}_2(\tilde{\epsilon})\} = {\mathbf P}(2\epsilon\tilde{\epsilon}).
\eeq{susy2regwj}
 If on  ($C^\infty (\Pi T^*{\cal L}M)$, $\{\,\,,\,\,\}$) there exist two generators which satisfy (\ref{susyalg1})
  and (\ref{susy2regwj}) then we say that there exists an $N=2$ supersymmetry. 

 By dimensional arguments, there is a unique ansatz 
  for the generator ${\mathbf Q}_2(\epsilon)$ on $\Pi T^*{\cal L}M$ which
  does not involve any dimensionful parameters
\beq
 {\mathbf Q}_2(\epsilon) = - \frac{1}{2} \int\limits_{S^{1,1}} d\sigma d\theta\,\,\epsilon
 ( 2 D\Phi^\rho S_\nu J^\nu_{\,\,\rho} + D\Phi^\nu D\Phi^\rho L_{\nu\rho} + 
 S_\nu S_\rho P^{\nu\rho}). 
\eeq{definchat}
 We can combine $D\Phi$ and $S$ into a single object 
\beq
\Lambda = \left ( \begin{array}{l}
                    D\Phi\\
                      S
\end{array} \right) ,
\eeq{definels}
 which can be thought of as a section of the pullback of  $X^*(\Pi (TM\oplus T^*M))$. The tensors in (\ref{definchat})
 can be combined into a single object
\beq
 {\cal J} = \left ( \begin{array}{ll}
                    - J & P\\
                     L &  J^t
\end{array} \right),
\eeq{definstardo}
 which is understood now as ${\cal J}: TM \oplus T^*M \rightarrow TM\oplus T^*M$.
 With this new notation we can rewrite (\ref{definchat}) as follows
\beq
  {\mathbf Q}_2(\epsilon) = - \frac{1}{2} \int\limits_{S^{1,1}} d\sigma d\theta\,\,\epsilon \langle
 \Lambda, {\cal J} \Lambda \rangle ,
\eeq{regska;}
   where $\langle\,\,,\,\,\rangle$ is understood as the induced pairing on $X^*(\Pi (TM\oplus T^*M))$. 
  The following proposition from \cite{Zabzine:2005qf} tells us when there exists $N=2$ supersymmetry.
 \begin{pro}
  $\Pi T^*{\cal L}M$ admits $N=2$ supersymmetry if and only
  if $M$ is a generalized complex manifold.
 \end{pro}
{\it Proof:}  We have to impose the algebra (\ref{susy2regwj}) on ${\mathbf Q}_2(\epsilon)$.
 The calculation of the second bracket is lengthy but straightforward and the corresponding coordinate 
 expressions are  given in \cite{Lindstrom:2004iw}. Therefore we give only the final 
 result of the calculation. Thus the algebra (\ref{susy2regwj}) satisfied if and only if 
\beq 
{\cal J}^2=-1_{2d},\,\,\,\,\,\,\,\,\,\,\,
 \Pi_{\mp} [\Pi_{\pm}(X+\eta), \Pi_{\pm}(Y+\eta)]_c=0,
 \eeq{definsgencomp}
 where $\Pi_\pm=\frac{1}{2}(1_{2d} \pm i {\cal J})$. Thus (\ref{definsgencomp}) together with the fact that 
  ${\cal J}$ (see (\ref{definstardo})) respects  the natural pairing (${\cal J}^t {\cal I} = - {\cal I} {\cal J}$)
 implies that ${\cal J}$ is a generalized complex structure.  $\Pi_\pm$ project to two 
  maximally isotropic involutive subbundles ${\mathbb L}$ and $\bar{\mathbb L}$ such that
$(T\oplus T^*)\otimes {\mathbb C} = {\mathbb L} \oplus \bar{\mathbb L}$.
  Thus we have shown that $\Pi T^*{\cal L}M$ admits $N=2$ supersymmetry if and only
  if $M$ is a generalized complex manifold. Our derivation is algebraic in nature 
   and does not depend on the details of  the model.   $\Box$
 
 The canonical transformations of $\Pi T^*{\cal L}M$ cannot change any brackets.
  Thus the canonical transformation corresponding to a b-transform (\ref{canonfr;as})
\beq
\left ( \begin{array}{l}
                    D\Phi\\
                      S
\end{array} \right)\,\,\,\rightarrow\,\,\,
\left ( \begin{array}{ll}
                    \,\,\,\,\,1 & 0\\
                   -b & 1
\end{array} \right)
\left ( \begin{array}{l}
                    D\Phi\\
                      S
\end{array} \right)
\eeq{howbtrahsdl}
 induces the following transformation of the generalized complex structure 
\beq 
{\cal J}_b = \left ( \begin{array}{ll}
                    1 & 0\\
                    b & 1
\end{array} \right) {\cal J} \left ( \begin{array}{ll}
                    \,\,\,\,\,1 & 0\\
                   -b & 1
\end{array} \right) 
\eeq{transofrmajdl}
 and thus gives rise to a new extended supersymmetry generator. 
 Therefore ${\cal J}_b$ is again the generalized complex
 structure. This is a physical explanation of the behavior of generalized complex 
 structure under $b$-transform.

Using $\delta_i (\epsilon) \bullet =\{ {\mathbf Q}_i(\epsilon), \bullet\}$ 
  we can write down the explicit form for the second
 supersymmetry transformations as follows
\beq
\delta_2(\epsilon) \Phi^\mu = i \epsilon D\Phi^\nu J^\mu_{\,\,\nu} -  i \epsilon S_\nu P^{\mu\nu}
\eeq{trsnakdf;}
\beq
\delta_2(\epsilon) S_\mu = i \epsilon D(S_\nu J^\nu_{\,\,\mu}) 
- \frac{i}{2} \epsilon S_\nu S_\rho P^{\nu\rho}_{\,\,\,\,,\mu}
 + i \epsilon D(D\Phi^\nu L_{\mu\nu}) + i \epsilon S_\nu D\Phi^\rho J^\nu_{\,\,\rho,\mu} -
 \frac{i}{2} \epsilon D\Phi^\nu D\Phi^\rho L_{\nu\rho,\mu}.
\eeq{tarsjS}
 Indeed it coincides with the supersymmetry transformation 
  analyzed in \cite{Lindstrom:2004iw}. 
  
    Also we could look for $N=2$ supersymmetry for $C^\infty (\Pi T^*{\cal L}M)$ with $\{\,\,,\,\,\}_H$.
     Indeed the result is exactly the same but now we have to have a twisted generalized complex
      manifold. 

 Another comment: We may  change the $N=2$ supersymmetry algebra
   (\ref{susyalg1}) and (\ref{susy2regwj}) slightly. Namely we can replace the last bracket in (\ref{susy2regwj}) by 
   \beq
    \{ {\mathbf Q}_2(\epsilon), {\mathbf Q}_2(\tilde{\epsilon})\} = -  {\mathbf P}(2\epsilon\tilde{\epsilon}).
   \eeq{djwlllw3930}
    This new algebra is sometimes called $N=2$ pseudo-supersymmetry. In this case we still 
     use the ansatz (\ref{definchat}) for ${\mathbf Q}_2$. However now we get
   \begin{pro}
  $\Pi T^*{\cal L}M$ admits $N=2$ pseudo-supersymmetry if and only
  if $M$ is a generalized product manifold.
 \end{pro}
   The proof of this statement is exactly the same as before. The only difference is that the condition 
    ${\cal J}^2=-1_{2d}$ get replaced by ${\cal J}^2=1_{2d}$.

\subsection{BRST interpretation}
\label{BRST}

Alternatively we can relate the generalized complex structure to an odd differential
  ${\mathbf s}$ on $C^\infty(\Pi T^*{\cal L}M)$ and thus we enter the realm of Hamiltonian 
   BRST formalism.  This formalism was developed to quantize theories with the first-class 
    constraints. 
  
  Indeed the supersymmetry generators (\ref{manisfetsssu})
 and (\ref{definchat}) can be thought of as odd transformations (by putting formally $\epsilon =1$)
 which square to the translation generator. Thus we can define the odd generator
\beq
{\mathbf q} = {\mathbf Q}_1(1) + i {\mathbf Q}_2(1) = - \int\limits_{S^{1,1}} d\sigma d\theta\,\,(S_\mu Q\Phi^\mu
 + i D\Phi^\rho S_\nu J^\nu_{\,\,\rho} + \frac{i}{2}D\Phi^\nu D\Phi^\rho L_{\nu\rho} + 
 \frac{i}{2} S_\nu S_\rho P^{\nu\rho}),
\eeq{nilpotenal}
 which is called the BRST generator.
  The odd generator ${\mathbf q}$ generates to the following transformation ${\mathbf s}$
\beq
{\mathbf s} \Phi^\mu = \{ {\mathbf q}, \Phi^\mu\}= Q\Phi^\mu + iD\Phi^\nu J^\mu_{\,\,\nu} - i S_\nu P^{\mu\nu} ,
\eeq{nilp11}
\beq
{\mathbf s} S_\mu = \{ {\mathbf q}, S_\mu \}= QS_\mu + i D(S_\nu J^\nu_{\,\,\mu}) 
- \frac{i}{2}  S_\nu S_\rho P^{\nu\rho}_{\,\,\,\,,\mu}
 + i D(D\Phi^\nu L_{\mu\nu}) + i S_\nu D\Phi^\rho J^\nu_{\,\,\rho,\mu} -
 \frac{i}{2}  D\Phi^\nu D\Phi^\rho L_{\nu\rho,\mu}, 
\eeq{nilp22}
 which is nilpotent due the properties  of manifest and nonmanifest supersymmetry trasnformations. 
 Thus ${\mathbf s}^2 =0$ if and only if ${\cal J}$ defined in (\ref{definstardo}) is a generalized
 complex structure. In doing the calculations one should remember that now 
 ${\mathbf s}$ is odd operation and whenever it passes through an odd object (e.g., $D$, $Q$ and
 $S$) there is extra minus. The existence of odd nilpotent operation (\ref{nilp11})-(\ref{nilp22}) is 
  typical for models with an $N=2$ supersymmetry algebra and corresponds to
    a topological twist of  the $N=2$ algebra.
    
    We can also repeat the argument for the $N=2$ pseudo-supersymmetry algebra and now 
     define the odd BRST generator as follows
     \beq
      {\mathbf q} = {\mathbf Q}_1(1) +  {\mathbf Q}_2(1).
     \eeq{neksk39939} 
  This  ${\mathbf q}$ generates an odd nilpotent symmetry if there exists a generalized product 
   structure. 
   
   We can equally well work with the twisted bracket $\{\,\,,\,\,\}_H$ and all results will be still valid
    provided that we insert the word "twisted" in appropriate places. We can summarize
     our discussion in the following proposition.
  
\begin{pro}
 The superPoisson algebra $C^\infty(\Pi T^*{\cal L}M)$ with $\{\,\,,\,\,\}$ ($\{\,\,,\,\,\}_H$) 
  admits odd derivation ${\mathbf s}$ if and only if there exists on $M$ either (twisted) generalized 
   complex or (twisted) generalized product structures. \\
    In other words the existence of an odd derivation ${\mathbf s}$ on $C^\infty(\Pi T^*{\cal L}M)$ 
     is related to real (complex) Lie bialgebroid structure on $TM \oplus T^*M$. 
\end{pro}

 The space $\Pi T^*{\cal L}M$ with odd nilpotent generator
  ${\mathbf q}$ can be interpreted as an extended phase space for a set of the first-class constraints 
   in $T^*LM$.  The appropriate linear combinations of $\rho$ and $\lambda$ are interpreted 
    as ghosts and antighosts.   The differential ${\mathbf s}$ on $C^\infty(\Pi T^*{\cal L}M)$
     induces the cohomology $H_{\mathbf s}^{\bullet}$ which is also a superPoisson algebra. 
  
  It is instructive to expand the transformations (\ref{nilp11})-(\ref{nilp22}) in components. 
   In particular if we look at the bosonic fixed points of the BRST action we arrive 
    at the following constraint
 $$ (1_{2d} + i {\cal J}) \left (\begin{array}{c}
 \d X\\
 p \end{array} \right) = 0 , $$
  which is exactly the same as the condition (\ref{defindlla78830}).  Thus we got the BRST 
   complex for the first-class constraints given by (\ref{firjsklllsl2929}). 
    These constraints correspond to TFTs as we have discussed,
   although the BRST complex above requires  more structure than just simply a (twisted) Dirac structure. 

\subsection{Generalized complex submanifolds}

 So far we have discussed the hamiltonian formalism for two dimensional field theory without 
  boundaries.  All previous discussion can be generalized to the case hamiltonian system
 with boundaries. 
 
 We start from the notion of a generalized submanifold.  Consider a manifold $M$ with
  a closed three form $H$ which specifies the Courant bracket. 
  \begin{Def}
   The data $(D, B)$ is called a generalized submanifold if $D$ is a submanifold of $M$ and
    $B \in \Omega^2(D)$ is a two-from on $D$ such that $H|_D = dB$. For any generalized
     submanifold we define a generalized tangent bundle 
     $$ \tau_D^B = \{ v +\xi \in TD \oplus T^*M|_D,\,\,\, \xi|_D = i_v B\} .$$
  \end{Def}  
  
  \begin{Exa}
   Consider a manifold $M$ with $H=0$, then any submanifold $D$ of $M$ is a generalized
    submanifold with $B=0$. The corresponding generalized tangent bundle is 
    $$ \tau^0_D = \{  v+ \xi \in TD \oplus N^*D \}$$
     with $N^*D$  being a conormal bundle of $D$.  Also we can consider $(D, B)$, 
      a submanifold with a closed two-form on it, $B\in \Omega^2(D)$, $dB=0$. Such 
      a pair $(D, B)$ is a generalized submanifold with generalized tangent bundle
        $$ \tau_D^B = e^B \tau_D^0,$$
         where the action of $e^B$ is defined in (\ref{actionofB}). 
  \end{Exa}
  
  The pure bosonic model is defined as follows. Instead of the loop space $LM$ we now 
   consider the path space
   $$ PM = \{ X: [0,1] \rightarrow M,\,\,\,X(0) \in D_0,\,\,X(1) \in D_1 \}$$
    where the end points are confined to prescribed submanifolds of $M$. The phase space 
     will be the cotangent bundle $T^*PM$ of path space. However to write down a symplectic 
      structure on $T^*PM$ we have to require that $D_0$ and $D_1$ give rise to 
       generalized submanifolds, $(D_0, B^0)$ and $(D_1, B^1)$, respectively.  Thus the symplectic
       structure on $T^*PM$ is
       $$ \omega = \int\limits_0^1 d \sigma \  \,\,(\delta X^\mu \wedge \delta p_\mu +  H_{\mu\nu\rho} \d X^\mu  \delta X^\nu \wedge \delta X^\rho ) +$$
    $$   + B^0_{\mu\nu}(X(0)) \delta X^\mu(0) \wedge \delta X^\nu (0)
        - B^1_{\mu\nu}(X(1)) \delta X^\mu(1) \wedge \delta X^\nu(1),$$
 where $\delta$ is de Rham differential on $T^*PM$. 
   It is crucial that $(D_0, B^0)$ and $(D_1, B^1)$ are generalized submanifolds
  for $\omega$ to be closed.   

 Next we have to introduce the super-version of $T^*PM$. This can be done in different
  ways. For example we can define the cotangent bundle $\Pi T^*{\cal P}M$ of superpath space 
   as the set of maps 
   $$\Pi T P \rightarrow \Pi T^*M$$ 
 with the appropriate boundary conditions which can be written as 
 $$ \Lambda (1) \in X^*(\Pi \tau_{D_1}^{B^1}) ,\,\,\,\,\,\,\,\,\,\,\,\,\,\,\,
  \Lambda (0) \in X^*(\Pi \tau_{D_0}^{B^0})$$
 with $\Lambda$ defined in (\ref{definels}).  These boundary conditions are motivated 
  by the cancellation of unwanted boundary terms in the calculations \cite{Zabzine:2005qf}.  
  
 Next we define a natural class of submanifold of a
  (twisted) generalized complex submanifold $M$. 

\begin{Def}
 A  generalized submanifold $(D,B)$ is called a generalized complex submanifold if 
  $\tau_D^B$ is stable under ${\cal J}$, i.e. if
  $$ {\cal J} \tau_D^B \subset \tau_D^B .$$ 
\end{Def}

Finally we would like to realize the $N=2$ supersymmetry algebra which has been discussed
 in previous subsections. The most of the analysis is completely identical to the previous 
  discussion. The novelty is the additional boundary terms in the calculations.  We present 
   the final result and skip all technicalities.

\begin{pro}
  $\Pi T^* {\cal P}M$ admits $N=2$ supersymmetry if and only if $M$ is a (twisted) generalized 
  complex manifold and $(D_i, B^i)$ are generalized complex submanifolds of $M$. 
 \end{pro}
 
 It is quite easy to generalize this result  to the real case when we talk about $N=2$ pseudo-supersymmetry. The correct notion would be a generalized product submanifold, i.e.
  such generalized submanifold $(D, B)$ when $\tau_D^B$ is stable under ${\cal R}$
   (see the definition \ref{defgenproduct} and the discussion afterwards). 
    This is quite straightforward and we will not discuss it here. 

\section*{LECTURE 3}
\addtocounter{section}{1}
\addcontentsline{toc}{section}{Lecture 3}

In this Lecture we review more advanced topics such as (twisted) generalized
 K\"ahler geometry and (twisted) generalized Calabi-Yau manifolds. In our presentation 
  we will be rather sketchy and give some of the statement without much elaboration. 
   We concentrate only on the complex case, although obviously there exists a real 
    version \cite{bonechi}.
  
  On physics side we would like to explain briefly that the generalized K\"ahler geometry 
   naturally arises when we specify the model, i.e. we choose a concrete Hamiltonian 
    in $C^\infty(\Pi T^*{\cal L}M)$, while the generalized Calabi-Yau conditions arise when 
     one tries to quantize this model. 

\subsection{Generalized K\"ahler manifolds}

$TM \oplus T^*M$ has a natural pairing $\langle\,\,,\,\,\rangle$. However one can introduce
 the analog of the usual positive definite metric.
 \begin{Def}\label{defmetric}
  A generalized metric is a subbundle $C_+ \subset TM \oplus T^*M$ of rank d ($\dim M =d$)
   on which the induced metric is positive definite 
 \end{Def}
  In other words we have splitting 
  $$ TM \oplus T^*M = C_+ \oplus C_-,$$
  such that there exists a positive metric on $TM \oplus T^*M$ given by 
  $$ \langle\,\,,\,\,\rangle|_{C_+} - \langle \,\,,\,\,\rangle|_{C_-} .$$
  Alternatively the splitting into $C_\pm$ can be described by an endomorphims
 $$ {\cal G} : TM \oplus T^*M \rightarrow TM \oplus T^*M,\,\,\,\,\,\,\,\,\,\,\,\,\,\,\,\,\,\,\,\,
 {\cal G}^2 =1,\,\,\,\,\,\,\,\,\,\,\,\,\,\,{\cal G}^t{\cal I} = {\cal I} {\cal G}, $$
  such that $\frac{1}{2}(1_{2d} \pm {\cal G})$ projects out $C_\pm$. 
   In order to write ${\cal G}$ explicitly we need the following proposition \cite{gualtieri}:
 \begin{pro}
  $C_\pm$ is the graph of $(b\pm g) : T \rightarrow T^*$ where 
   $g$ is Riemannian metric and $b$ is two form.  
 \end{pro}
 As a result ${\cal G}$ is given by 
 \beq
  {\cal G} = \left (\begin{array}{cc}
  1 & 0 \\
  b & 1 \end{array} \right ) \left (\begin{array}{cc}
  0 & g^{-1} \\
  g & 0 \end{array} \right ) 
 \left ( \begin{array}{cc} 
 1 & 0 \\
 - b & 1 \end{array} \right )
  =\left ( \begin{array}{cc}
   - g^{-1}b & g^{-1} \\
    g - b g^{-1} b & b g^{-1} \end{array} \right) .
 \eeq{defkf2wp2000}
 Thus the standard metric $g$ together with the two-form $b$ give rise to a generalized metric 
  as in the definition \ref{defmetric}. 
 
 Now we can define the following interesting construction. 
 \begin{Def}\label{defGKM}
  A (twisted) generalized K\"ahler structure is a pair ${\cal J}_1$, ${\cal J}_2$ of commuting (twisted)
   generalized complex structures such that ${\cal G} = - {\cal J}_1 {\cal J}_2$ is a positive definite metric
    (generalized metric) on $TM \oplus T^*M$. 
 \end{Def} 
  Indeed this is the generalization of the K\"ahler geometry as can been seen from 
   the following example.
 \begin{Exa}\label{exakahler}
  A K\"ahler manifold is a complex hermitian  manifold $(J, g)$ with a closed
   K\"ahler form $\omega = gJ$.  A K\"ahler manifold is an example of a generalized
    K\"ahler manifold where ${\cal J}_1$ is given by example \ref{exa123} and ${\cal J}_2$
     by  example \ref{exa1234}.  Since the corresponding symplectic structure $\omega$ is 
      a K\"ahler form,  two generalized complex structures commute and their product is 
      $$ {\cal G} = - {\cal J}_1 {\cal J}_2 = \left ( \begin{array}{cc}
      0 & g^{-1} \\
      g & 0 
      \end{array} \right ) .$$
       This example justifies the name, a generalized K\"ahler geometry. 
 \end{Exa}
  For (twisted) generalized K\"ahler manifold there are the following decompositions of 
   complexified tangent and cotangent bundle
  $$ (TM \oplus T^*M) \otimes \mathbb{C} = {\mathbb L}_1 \oplus \bar{\mathbb L}_1 = {\mathbb L}_2 \oplus \bar{\mathbb L}_2,$$
   where the first decomposition corresponds to ${\cal J}_1$ and second to ${\cal J}_2$.
   Since $[{\cal J}_1, {\cal J}_2]=0$ we can do both decompositions simultaneously 
   $$ (TM \oplus T^*M) \otimes \mathbb{C} = {\mathbb L}_1^+ \oplus {\mathbb L}_1^- \oplus \bar{\mathbb L}_1^+ \oplus \bar{\mathbb L}^-_1,$$
 where the space ${\mathbb L}_1$ ($+i$-egeinbundle of ${\cal J}_1$) can be decomposed into ${\mathbb L}_1^\pm$, 
  $\pm i$-egeinbundle of ${\cal J}_2$.  In its turn the generalized metric subbundles are defined as
    $$ C_\pm \otimes \mathbb{C} =  {\mathbb L}_1^\pm \otimes \bar{\mathbb L}_1^\pm . $$

  One may wonder if there exists an alternative geometrical description for a (twisted)
   generalized K\"ahler manifolds. Indeed there is one.
 
 \begin{Def}
  The Gates-Hull-Ro\v cek geometry is the following geometrical data: two complex structures $J_\pm$, 
   metric $g$ and closed three form $H$ which satisfy
    $$J_\pm^t g J_\pm = g$$
    $$\nabla^{(\pm)} J_\pm =0$$
     with the connections defined as $\Gamma^{(\pm)} = \Gamma \pm g^{-1} H$, where 
      $\Gamma$ is a Levi-Civita connection for $g$. 
 \end{Def}
  This geometry was originally derived by looking at the general $N=(2,2)$ supersymmetric 
   sigma model \cite{Gates:1984nk}.  In \cite{gualtieri} the equivalence of these two seemingly 
    unrelated descriptions has been proven. 
 
 \begin{pro}
  The Gates-Hull-Ro\v{c}ek geometry is equivalent to a twisted generalized K\"ahler geometry. 
 \end{pro}
 
 As we have discussed briefly a generalized complex manifold locally looks like a product 
  of symplectic and complex manifolds. The local structure of (twisted) generalized K\"ahler 
   manifolds is somewhat involved.
Namely the local structure is given by the set of symplectic foliations arising from 
  two real Poisson structures \cite{Lyakhovich:2002kc}
  and holomorphic Poisson structure \cite{hitchinP}. Moreover one can show that in analogy with 
   K\"ahler geometry there exists a generalized K\"ahler potential which encodes all local geometry 
    in terms of a single function \cite{Lindstrom:2005zr}. 

\subsection{$N=(2,2)$ sigma model}

In the previous Lecture we have discussed the relation between (twisted) generalized complex
 geometry and $N=2$ supersymmetry algebra on $\Pi T^*{\cal L}M$. Our discussion has been 
  model independent. A choice of concrete model corresponds to a choice of Hamiltonian 
   function ${\cal H}(a) \in C^\infty (\Pi T^*{\cal L}M)$  which generates a time evolution of a system. 
    Then the natural question to ask if the model is invariant under the $N=2$ supersymmetry,
     namely 
     \beq
      \{ {\mathbf Q}_2(\epsilon), {\cal H}(a)\} =0 ,
     \eeq{brajskw9930}
    where ${\mathbf Q}_2(\epsilon)$ is defined in (\ref{definchat}) with the corresponding (twisted)
    generalized complex structure ${\cal J}$. 
   
   To be concrete we can choose the Hamiltonian which corresponds to $N=(2,2)$ sigma model 
    used by Gates, Hull and Ro\v{c}ek in \cite{Gates:1984nk}
   $$  {\cal H}(a) = \frac{1}{2} \int d\sigma\,d\theta\,\,a \left ( i \d\phi^\mu D\phi^\nu g_{\mu\nu}
 + S_\mu D S_\nu g^{\mu\nu} + S_\sigma D\phi^\nu S_\gamma g^{\lambda\gamma}
  \Gamma^\sigma_{\,\,\,\,\nu\lambda}  - \right .$$
  \beq
\left .  - \frac{1}{3} H^{\mu\nu\rho} S_\mu S_\nu S_\rho 
   + D\phi^\mu D\phi^\nu S_\rho H_{\mu\nu}^{\,\,\,\,\,\,\rho}  \right),
  \eeq{fullhamwithevetey}
 where $a$ is just an even test function.   This Hamiltonian has been derived in \cite{Bredthauer:2006hf}. 
  This Hamiltonian is invariant under the $N=2$ supersupersymmetry if 
  $$ {\cal J}_1= {\cal J},\,\,\,\,\,\,\,\,\,\,\,\,\,\,\,{\cal J}_2 = {\cal J} G$$
 is a (twisted) generalized K\"ahler structure, see the definition \ref{defGKM}.  For the Hamiltonian 
  (\ref{fullhamwithevetey}) ${\cal G}$ is defined by (\ref{defkf2wp2000}) by $g$ 
   and $b=0$, $H$ corresponds 
   the closed three-form which is used in the definition of the twisted Courant bracket. 
    Indeed on  a (twisted) generalized K\"ahler manifold ${\cal H}$ is invariant under 
     supersymmetries corresponding to both (twisted) generalized complex structures, ${\cal J}_1$
      and ${\cal J}_2$. 
   
   Also the Hamiltonian (\ref{fullhamwithevetey}) can be interpreted in the context of TFTs.
     Namely ${\cal H}$  is the gauge fixed Hamiltonian for the TFT we have discussed in subsection
      \ref{BRST} with ${\mathbf s}$ being the BRST-transformations defined in (\ref{nilp11})-(\ref{nilp22}). 
     The Hamiltonian (\ref{fullhamwithevetey}) is BRST-exact
    $$ {\cal H} = {\mathbf s}\left (  \frac{i}{4}\int d\sigma d\theta\,\,\langle \Lambda, {\cal J} {\cal G} \Lambda 
    \rangle \right )  =   {\mathbf s}\left (  \frac{i}{4}\int d\sigma d\theta\,\,\langle \Lambda, {\cal J}_2 \Lambda 
    \rangle \right )    .$$
     Moreover the translation operator ${\mathbf P}$ is given by
     $$ {\mathbf P} = {\mathbf s} \left (  \frac{i}{4}\int d\sigma d\theta\,\,\langle \Lambda, {\cal J} \Lambda 
    \rangle \right ) =  {\mathbf s} \left (  \frac{i}{4}\int d\sigma d\theta\,\,\langle \Lambda, {\cal J}_1  \Lambda 
    \rangle \right ). $$
     The $N=(2,2)$ theory (\ref{fullhamwithevetey}) is invariant under two extended
      supersymmetries associated to generalized complex structures, ${\cal J}_1$ and ${\cal J}_2$. 
       Thus there are two possible BRST symmetries and correspondingly two TFTs associated 
        either to ${\cal J}_1$ or to ${\cal J}_2$.  In the literature these two TFTs are called either 
         A or B topological twists of the $N=(2,2)$ supersymmetric theory. 
    
    Indeed one can choose a different Hamiltonian function on $\Pi T^* {\cal L}M$ and 
     arrive to  different geometries which involve the generalized complex structure, e.g.
      see \cite{Calvo:2005ww}.

\subsection{Generalized Calabi-Yau manifolds}

In this subsection we define the notion of generalized Calabi-Yau manifold. To do this we have to
introduce  a few new concepts.  

We can define the action of a section $(v +\xi) \in \Gamma (TM \oplus T^*M)$
on  a differential form $\phi \in \Omega(M)= \wedge^{\bullet} T^*M$ 
$$ (v + \xi ) \cdot \phi \equiv i_v \rho + \xi \wedge \phi.$$
 Using this action we arrive at the following identity
$$\{ A, B\}_+\cdot \phi \equiv A\cdot (B\cdot \phi) + B \cdot (A \cdot \phi) = 2 \langle A, B\rangle \phi,$$
 which gives us the representation of Clifford algebra, $Cl(TM\oplus T^*M)$, on the differential forms. 
Thus we can view differential forms as spinors for $TM\oplus T^*M$ and moreover there are
  no topological obstructions for their existence.
   In further discussion we refer to a differential form as a spinor.
   
The   chirality decomposition for spinors corresponds to decomposing forms into even and 
 odd degrees, 
  $$\Omega(M)= \wedge^{\bullet} T^*M = \wedge^{even} T^*M \oplus \wedge^{odd}T^* M.$$
   We would like to stress that in all present discussion we do not consider a form of a definite degree, 
 but   we may consider a sum of the forms of different degrees. 
   Also on $\Omega(M)$ there exists a $Spin(d,d)$-invariant bilinear form $(\,\,,\,\,)$,  
  \beq
  (\phi, r) = [\phi \wedge \sigma (r)] |_{top},
  \eeq{definpairing}
   where $\phi, r  \in \Omega(M)$ and $\sigma$ is anti-automorphism which reverses the wedge product.
    In the formula (\ref{definpairing}) $[...]|_{top}$ stands for the projection to the top form.

\begin{Def}
 For any form $\phi \in \Omega(M)$ we define a null space 
 $$ {\mathbb L}_\phi = \{ A\in \Gamma(TM \oplus T^*M), A\cdot \phi =0\}$$
 \end{Def}
  Indeed the null space ${\mathbb L}_\phi$ is isotropic since
  $$ 2 \langle A, B \rangle \phi = A\cdot (B\cdot \phi) + B \cdot (A \cdot \phi ) =0$$

\begin{Def}
 A spinor $\phi \in \Omega(M)$ is called pure when ${\mathbb L}_\phi$ is a maximally isotropic 
  subbundle of $TM \oplus T^*M$ (or its complexification). 
\end{Def}

\begin{pro}
 ${\mathbb L}_\phi$ and ${\mathbb L}_r$ satisfy ${\mathbb L}_\phi \cap {\mathbb L}_r =0$ if and only if
 $$ (\phi, r ) \neq 0,$$
  where $(\,\,,\,\,)$ is bilinear form defined in (\ref{definpairing}). 
\end{pro}
Obviously  all this can be complexified.
  
  If we take a pure spinor $\phi$ on $(TM \oplus T^*M) \otimes {\mathbb C}$ such that 
   $(\phi, \bar{\phi}) \neq 0$ then the complexified  tangent plus cotangent bundle 
    can be decomposed into the corresponding null spaces 
   $$ (TM \oplus T^*M) \otimes {\mathbb C} = {\mathbb L}_\phi \oplus {\mathbb L}_{\bar{\phi}} = {\mathbb L}_\phi \oplus 
    \bar{\mathbb L}_\phi .$$
 Therefore we have an almost generalized complex structure. 
 
 The following definition is due to Hitchin \cite{hitchinCY}. However we follow the 
  terminology proposed in \cite{Li:2005tz}.
 \begin{Def}\label{defCY}
  A weak generalized Calabi-Yau manifold is a manifold with a pure spinor $\phi$
   such that $(\phi, \bar{\phi}) \neq 0$ and $d\phi =0$. 
 \end{Def}
 
 A weak generalized Calabi-Yau manifold is generalized complex manifold since ${\mathbb L}_\phi$ 
  and ${\mathbb L}_{\bar{\phi}}$ are complex Dirac structures. The condition $d\phi =0$ implies
   the involutivity of ${\mathbb L}_\phi$.  There is also a twisted weak generalized Calabi-Yau manifold
    where in the definition \ref{defCY} the condition $d\phi =0$ is replaced by the condition 
     $d\phi + H \wedge \phi =0$. The twisted weak generalized Calabi-Yau manifold is a twisted 
      generalized complex manifold.
      
 \begin{Exa}
  In Example \ref{exa1234} we have considered the symplectic manifold and have argued 
   that there exists the generalized complex structure. Indeed a symplectic manifold 
    is a weak generalized Calabi-Yau manifold with a pure spinor given by 
     $$ \phi = e^{i \omega} = 1 + i \omega + \frac{i^2}{2} \omega \wedge \omega + ... + \frac{i^n}{n!}
     \omega \wedge ...
      \wedge \omega$$
       with the last term on the right hand side corresponding to a top form. 
 \end{Exa}

\begin{Exa}
 A  complex manifold is a generalized complex manifold, see example \ref{exa123}. 
  However it is not a weak generalized Calabi-Yau manifold automatically. We have to 
   require the existence of a closed holomorphic volume form (the same as a closed holomorphic top 
    form nowhere vanishing) 
   $$ \phi = \Omega^{(n,0)}$$
    which corresponds to a pure spinor. 
\end{Exa}

We would like to stress that  any weak generalized Calabi-Yau manifold is a generalized complex manifold, but not vice versa.

\begin{Def}
\label{genCYa}
 A generalized Calabi-Yau manifold is a manifold with two closed pure spinors, 
  $\phi_1$ and $\phi_2$ such that 
 $$ (\phi_1 , \bar{\phi}_1) = c (\phi_2, \bar{\phi}_2) \neq 0$$
  and they give rise to a generalized K\"ahler structure.
\end{Def}
 Also we can define a twisted Calabi-Yau manifold where in the above definitions
  the spinors satisfy $(d+H \wedge) \phi_i =0$ and they give rise to a twisted generalized 
   K\"ahler geometry. 
   
   \begin{Def}
     A standard Calabi-Yau manifold is a K\"ahler manifold (see the example \ref{exakahler})
      with a closed holomorphic volume form $\Omega^{(n,0)}$.  This gives us
     an example of generalized Calabi-Yau manifold with $\phi_1 = e^{i\omega}$ and 
      $\phi_2 = \Omega^{(n,0)}$.
       \end{Def}

\subsection{Quantum $N=(2,2)$ sigma model}

In this subsection we would like to discuss very briefly the quantization of $N=(2,2)$
 sigma model given by (\ref{fullhamwithevetey}) and its corresponding TFTs cousins.
  In all generality this problem  is a hard one and remains unresolved. 
     Although it is always simpler to quantize TFTs. 
   However by now we understand that for a $N=(2,2)$ sigma model 
    to make sense at the quantum level we have to require  the generalized Calabi-Yau conditions. 
    We are going briefly sketch the argument which was presented essentially in 
 \cite{Kapustin:2004gv}.  
 
  We start our discussion from the TFT associated to a generalized complex structure. 
 It is not simple to quantize a theory in all generality. However it is convenient to look first 
  at the semiclassical approximation. It means that we can ignore $\sigma$ dependence 
   and all loops collapse to a point on $M$. Thus we replace\footnote{$\Pi T^*{\cal L}M$ collapses 
    to $T^*(\Pi TM)$ since $\sigma$ dependence disappear but $\theta$-dependence is still there.
     See \cite{rot} for the detailed discussion of $T^*(\Pi TM)$ and related matters.} $\Pi T^*{\cal L}M$ by  $T^*(\Pi TM) \approx
    T (\Pi T^*M)$
    and try to quantize this simpler theory. In particular we have to interpret the generator ${\mathbf q}$
     (\ref{nilpotenal}) restricted to $T^*(\Pi TM)$.  For this we have to expand the generator ${\mathbf q}$
     (\ref{nilpotenal}) in components and drop all terms which contain the derivatives with 
      respect to $\sigma$. Moreover it is useful to rotate  the odd basis $(\lambda^\mu, \rho_\mu) \in\Gamma(\Pi(TM  \oplus T^*M))$   to a new one $(\xi^A, \bar{\xi}_A) \in \Gamma(\Pi ({\mathbb L} \oplus \bar{\mathbb L}))$  which is adopted 
       to $\pm i$-eigenbundles of ${\cal J}$. $\xi_A$ correspond to ghosts and $\bar{\xi}^A$ to antighosts.
       After these manipulations ${\mathbf q}$ can be 
        written as follows
        \beq
        {\mathbf q} \sim p_\mu \rho^{\mu A}(X) \xi_A   +  f^{AB}_{\,\,\,\,\,\,C}(X) \xi_A \xi_B \bar{\xi}^C,
        \eeq{defjsoo399400}
        where we have ignored the irrelevant overall numerical factor.  Now in new 
         odd basis our phase space is $T^*(\Pi {\mathbb L}) \approx T^*(\Pi \bar{\mathbb L})$. 
        We remember that ${\mathbb L}$
         is a Lie algebroid and thus $\rho^{\mu A}(X)$ and $f^{AB}_{\,\,\,\,\,\,C}(X)$ are the anchor
          map and structure constants defined in subsection \ref{subalgebroid}. This reduced 
           ${\mathbf q}$ acts naturally on $\wedge^{\bullet} \bar{\mathbb L}= C^\infty(\Pi {\mathbb L})$ 
            and gives rise to so-called 
            Lie algebroid cohomology $H(d_L)$.  In TFT we would associate the set of local observables 
             to the elements of $H(d_L)$. 
             
        Also in any quantum field theory we have to build a Hilbert space of states. If we regard
     $(\lambda^\mu, \rho_\mu) $ as a set  of creation and annihilation operators then the corresponding 
      Fock space will be given by  $\Omega(M)$. Alternatively we could choose  $(\xi^A, \bar{\xi}_A)$
         as a set of creation and annihilation operators. This choice would induce the natural 
          grading
          $$ \Omega (M) = U_0 \oplus ( \bar{\mathbb L} \cdot U_0 )\oplus (\wedge^2 \bar{\mathbb L} \cdot U_0 )\oplus ...
          \oplus (\wedge^d {\bar{\mathbb L}} \cdot U_0),$$
 where $U_0$ is a vacuum state over which we build the Fock space. Mathematically we could 
  choose $U_0$ to be a pure spinor line (i.e., we use the existence of pure spinor only locally). 
   The operator ${\mathbf q}$ now acts on $\Omega(M)$ and it induces another cohomohology  $H(\bar{\d})$, 
    which corresponds to a Hilbert space.
   
   Next we cite the following theorem without a proof.
   \begin{pro}
    For a (twisted) weak generalized Calabi-Yau manifold we have an isomorphism of two 
     cohomologies
     $$H(d_L) \sim H(\bar{\d})$$
   \end{pro}

 For the  TFT the isomorphism of these two cohomologies is interpreted as operator-state correspondence, for each local observable we can associate a state in a Hilbert space and vice versa.  Thus if we want to have the operator-state correspondence the corresponding TFT should be 
  defined over a (twisted) 
   weak Calabi-Yau manifold. Indeed there are more  interesting structures in this
   TFT about which we do not have time to talk, see
  \cite{Li:2005tz}, \cite{Pestun:2006rj}. 
  
  Let us finish with a few comments about the $N=(2,2)$ sigma model. The 
  above analysis of states in TFT
   corresponds to analysis of the ground states in the $N=(2,2)$ 
  sigma model.  At the level of ground states there should be also the operator-state correspondence
   and thus we have to require a (twisted)
     weak Calabi-Yau structure for both ${\cal J}_1$ and ${\cal J}_2$. 
    Since ${\cal J}_1$ and ${\cal J}_2$ correspond to a (twisted) generalized K\"ahler structure 
     we arrive to the definition of (twisted) generalized Calabi-Yau manifold. Thus we can conclude 
      with the following proposition.
      
      \begin{pro}
       The quantum $N=(2,2)$ sigma model requires $M$ to be a (twisted) generalized 
        Calabi-Yau manifold, see the definition \ref{genCYa}.
      \end{pro}
  
\subsection{Summary}

 In these lecture notes we made an attempt to introduce the concepts of the generalized geometry 
  and its relevance for the string theory.  We concentrated our attention on the Hamiltonian 
   approach to the world-sheet theory. Due to lack of time we did not discuss other important issues
    within the world-sheet theory, see the contribution \cite{Lindstrom:2006ee} to the same volume 
     for a review and the references.

 Another topic which we did not touch at all concerns the space-time aspects of the generalized 
  geometry, see  \cite{Grana:2005jc} for the review and references. Eventually the world-sheet 
   point of view is ultimately related to the space-time aspects of the problem.
   
   Finally we have to stress that presently the subject is actively developing and 
    there are still many unresolved problems. 

\noindent{\bf Acknowledgement}:  I am grateful to the organizers of the Winter School
 for the invitation and for the warm hospitality during my stay in Czech Republic. 
  I thank Francesco Bonechi, Ulf Lindstr\"om, Pierre Vanhove and Rikard von Unge
    for the reading and commenting on these lecture notes.
The research  is supported by  by VR-grant 621-2004-3177.

\end{document}